\documentclass[manuscript]{acmart}

\usepackage{tabularx}

\AtBeginDocument{
  \providecommand\BibTeX{{
    \normalfont B\kern-0.5em{\scshape i\kern-0.25em b}\kern-0.8em\TeX}}}

\copyrightyear{2020}
\acmYear{2020}
\setcopyright{acmlicensed}
\acmConference[PDC '20: Vol. 2]{Proceedings of the 16th Participatory Design Conference 2020- Participation(s) Otherwise - Vol. 2}{June 15--20, 2020}{Manizales, Colombia}
\acmBooktitle{Proceedings of the 16th Participatory Design Conference 2020- Participation(s) Otherwise - Vol. 2 (PDC '20: Vol. 2), June 15--20, 2020, Manizales, Colombia}
\acmPrice{15.00}
\acmDOI{10.1145/3384772.3385143}
\acmISBN{978-1-4503-7606-8/20/06}

\begin{document}

\title[Scaling Participation]{Scaling Participation - What Does the Concept of Managed Communities Offer for Participatory Design?}


\author{Stefan Hochwarter}
\email{stefan.hochwarter@ntnu.no}
\affiliation{%
  \institution{Norwegian University for Science and Technology}
  \streetaddress{Høgskoleringen 1}
  \city{Trondheim}
  \state{Norway}
  \postcode{7491}
}

\author{Babak A. Farshchian}
\email{babak.farshchian@ntnu.no}
\affiliation{%
  \institution{Norwegian University for Science and Technology}
  \streetaddress{Høgskoleringen 1}
  \city{Trondheim}
  \state{Norway}
  \postcode{7491}
}


\begin{abstract}
    This paper investigates mechanisms for scaling participation in participatory design (PD). Specifically, the paper focuses on \textit{managed communities}, one strategy of generification work. We first give a brief introduction on the issue of scaling in PD, followed by exploring the strategy of managed communities in PD. This exploration is underlined by an ongoing case study in the healthcare sector, and we propose solutions to observed challenges. The paper ends with a critical reflection on the possibilities managed communities offer for PD. Managed communities have much to offer beyond mere generification work for large-scale information systems, but we need to pay attention to core PD values that are in danger of being sidelined in the process.
\end{abstract}



\newcommand{\citneeded}{[citation needed]}

\maketitle

\section{Introduction}

In a recent article, Bødker and Kyng~\cite{bodkerParticipatoryDesignThat2018} propose an agenda for Participatory Design (PD) that Matters. One point they make is the need to scale PD to make it relevant for the design of large IT systems and infrastructures. The issue of scaling is not new in PD (see e.g.~\cite{simonsenParticipativeDesignChallenges2008}), but the democratic implications of (not) being able to support PD on larger scales, especially in modern infrastructures, are rarely addressed: "In our view, the centralization of (the major commercial platforms on) the Internet, big data, and large-scale infrastructuring challenge the core democratic ideals of PD"~\cite[p.8]{bodkerParticipatoryDesignThat2018}.

PD has historically focused on the democratisation of practices and equalisation of power relations~\cite{bodkerParticipatoryDesignThat2018,kensingHeritageHavingSay2012}. These values need to be considered when discussing the scaling of PD. Our research interest is in investigating scaling mechanisms with these fundamental PD values in mind.

In this paper we aim to explore a specific strategy, i.e. \textit{managed communities} by Pollock et al.~\cite{pollockGlobalSoftwareIts2007} as a scaling strategy for design work in PD. Software vendors actively engage managed communities -- i.e. large groups of users from multiple current or future customer organisations -- in continuous (re)design of their products. While managed communities in Pollock et al.'s original paper were described from a top-down managerial perspective, we believe the concept and its processes are valuable for discussing the scaling of PD as well. We investigate the managed communities strategy with insights from an ongoing case study of a software company that engages its customers and users in large yearly user conferences. In this early stage of our case study we look into how these conferences are organised, and what role they play as managed communities in PD. For this we combine Pollock et al.'s concept of managed communities with a framework proposed by Borum and Enderud~\cite{borumKonflikterOrganisationerBelyst1981}. According to this framework, the suitability of user conferences as a PD tool depends on their agenda, participants, scope, and resources (such as time) available to users. Combining Pollock et al.'s concept of managed communities with the framework of Borum and Enderud, the contribution of our paper to PD research is a discussion of whether managed communities, originally described as a management tool, can be modified to become a scaling tool for PD.

In the rest of this paper we first introduce some relevant related work on managed communities. We then present our case and our preliminary findings before we conclude with a discussion.

\section{Managed communities and PD}

\begin{figure}[h]
  \centering
  \includegraphics[width=0.6\linewidth]{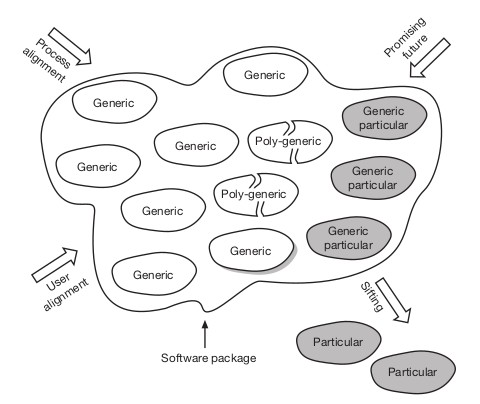}
  \caption{Generic solution as a 'black-blob'. Figure is from~\cite{pollockGlobalSoftwareIts2007}}.
  \Description{Generic solution as a 'black-blob'}
  \label{fig:black-blob}
\end{figure}

Achieving scale while at the same time not losing focus of core values has drawn recent attention in the PD community~\cite{bodkerParticipatoryDesignThat2018,wagnerCriticalReflectionsParticipation2018,rolandPlatformArchitecturesLargescale2017}. Within this line of research, our inspiration for this paper is from Pollock et al.~\cite{pollockGlobalSoftwareIts2007} exploring the following paradox: scholars often emphasize the importance of organisational contexts to the design of software systems, while at the same time there exist systems that work across these diverse contexts, even on a global scale. Pollock et al. use the term \textit{generification work} to denote those strategies and processes that allow systems to be designed in such a way that they can travel across different organisational settings. User involvement plays a central role in generification work. A mechanism that companies use to keep users and customers in the loop is called \textit{managed communities}. As part of the generification, "the translation from a particular to a generic technology corresponds to a shift from a few isolated users to a larger extended ‘community’"~\cite[p.261]{pollockGlobalSoftwareIts2007}. During generification work, where users play an active role, mutual learning (\textit{witnessing}) is observed as a positive by-product. Users become aware of the complexity of designing and implementing large-scale systems while the supplier of the system observes the differences and similarities across the settings. Vendors of systems for large markets often leave their systems open for customisation by the users, allowing particular requirements on a per-user basis. Further, users can potentially be empowered to participate, find new functionalities and help improve the product~\cite{vonhippelDemocratizingInnovationEvolving2005}. Figure~\ref{fig:black-blob} shows how generification shapes a system. In short, the \textit{generic} blob in the figure denotes what is common across almost all user groups, the \textit{poly-generic} is similar functionality varying little between user groups, the \textit{generic particular} can be functionality specific for a user group but included in the core product due to strategic reasons, and the \textit{particular} is not included in the system and left to the user groups to implement (or not).

 Instruments similar to managed communities are reported by other researchers as well. One such example are the \textit{academies} in the Health Information System Programme (HISP)~\cite{braaHealthInformationSystems2012}, an action research project that explicitly focuses on capacity building. HISP hosts academies at different expertise levels to include their users in the design process, facilitate requirement elicitation and provide a learning venue. Another example is reported by Eaton et al.~\cite{eatonDistributedTuningBoundary2015}, where they investigate online communities of third-party application developers who use Apple's AppStore platform and its APIs (Application Programming Interfaces). They show how Apple tries -- and sometimes fails -- to manage its communities of developers. Both cases also demonstrate PD in the context of the increasingly dominant platform model for software. The platform architecture supports particular requirements for specific users in its periphery, while the platform core holds generic requirements across different user groups.

 The concept of generification work is described by Pollock et al. from a managerial perspective. We want to investigate whether it can also be applied with the core values of PD in mind. Consequently, the users' role needs to shift from being informed to actively participating in decision-making. The term \textit{managed} communities implies that they are managed. We need to ask: Managed, by whom and with what interest? If the power gap between the organisers and the participants is too big, managed communities will not support shared decision-making and rather strengthen the actor in power~\cite{bratenModelMonopolyCommunication1973}.

Following the model of Borum and Enderud (1981)~\cite{borumKonflikterOrganisationerBelyst1981}, organisers of managed communities need to be cautious regarding the four mechanisms that are influencing this power relation: (i) agenda control, (ii) participants, (iii) scope, and (iv) resources. A managed community comes with an \textbf{agenda}, commonly set by the company or organisation behind the development of the system. The agenda is designed with a certain outcome in mind, for example finding new requirements, smoothing requirements or gathering users' feedback on newly introduced functionalities. Large-scale communities have \textbf{participants} and users spread geographically. Selecting and inviting specific users shape the context in which the managed community and consequently the shaping of the system will take place. The \textbf{scope} of the community is also related to its agenda and participants, and how much freedom is given to them to define this scope. Connected to all three mechanisms used to exercise power is the mechanism of \textbf{resources}. Not all users who want to participate to the community have the resources required. For instance, time is a resource often scarcely available and time-intensive peak times vary across domains and regions.

\section{The empirical case and methods}

The ongoing case study (started in March 2019) is part of an emerging multiple, embedded case study research design~\cite{yinCaseStudyResearch2017} investigating the design, implementation and use of assistive technologies in health care. The case takes place at a health IT company in one of Norway's largest municipalities. In the following, we will call this company \textit{HealthSoftWare}. The municipality partners with HealthSoftWare to test, implement and deliver assistive technologies to the population, as specified by the Norwegian welfare technology program~\cite{helsedirektoratetVelferdsteknologiFagrapportOm2012}.  In the following we will shortly describe the history of the company and the biography of the artefact that is of interest~\cite{pollockSoftwareOrganizationsBiography2012}.

\subsection{The organisational context}

HealthSoftWare is specialised in providing self-reporting and self-management systems for healthcare organisations and clinical studies. The product idea was born in 2007 in the IT department of a private clinic. Initially, it focused on online patient-managed questionnaires for pain management and mental health.

Professional development of the product started in 2009 and a business plan was put in place. While the previous version was driven by internal needs, although concepts were perceived as good and useful, the use of the system had remained low due to missing reimbursement possibilities. The system was redesigned to create volume and overcome reimbursement issues. From 2009 to 2014 the HealthSoftWare product changed from a solution to a platform for managing online healthcare-related questionnaires. The goal was to eventually target an international market, hence lock-in situations by exclusively supporting national standards were avoided. In addition, HealthSoftWare followed a modular, flexible architecture which allowed them to quickly react to market requirements.

HealthSoftWare positioned themselves as an intermediary between their customers among healthcare service providers and researchers, and the providers of the standardised questionnaires. They did not pursue to become exclusive right holders to these questionnaires but aimed to provide a platform for the providers to reach a large audience -- i.e. a platform model. The system was designed to be open, both partners and customers had access to the API for integration with their existing systems. By 2014, a studio module allowed customers to design and use their own questionnaires on the platform. User generated content was not any more isolated to a single instance but could travel across systems and organisations.

In their major next version, HealthSoftWare focused on scalability and user-based customisation of the system. More advanced communication was introduced, such as secure chats with the ability to add attachments or the implementation of real time communication. Further, the user interface for patients was designed to be mobile first and responsive, in the light of increased use of tablets and smartphones.

In 2018, one of Norway's largest municipalities was looking for a solution to implement self-reporting among patients with chronic conditions. By this time, HealthSoftWare fulfilled the major requirements and adding new devices and sensors to their system was supported through their API. They won the tender and extended their existing system to fulfil the project requirements.

\subsection{Methods for data collection}

Data presented here was mainly collected by participant observation (during a user conference) and semi-structured interviews (1 interview with CEO and 3 interviews with the CTO). Collection of data took place between March and October 2019. The interviews were held in English and the information received at the user conference was in Norwegian. In addition to the observation, additional documents such as presentation slides and brochures from the user conference were collected and analysed.

In our data collection we have focused on two instruments for managing communities, i.e. user conferences and training courses. We will describe these in the next section.

\section{Findings}

HealthSoftWare organises regular yearly user conferences called \textit{HealthSoftWare Conference}. These are spread over two days, each day covering one major part of the conference: (i) presentations mainly provided by user groups, i.e. the \textit{conference}, and (ii) courses provided by HealthSoftWare, called \textit{course and inspiration}. Table \ref{tab:themes} lists the theme of recent conferences.

\begin{table}
  \caption{Themes of HealthSoftWare conferences}
  \label{tab:themes}
  \begin{tabular}{ll}
    \toprule
    Year&Theme\\
    \midrule
    2016 & Digital self-reporting and self-management \\
    2017 & Towards digital patient participation \\
    2018 & no conference due to moving conference to spring \\
    2019 & Digital healthcare services \\
    2020 & The value of digital healthcare services \\
  \bottomrule
\end{tabular}
\end{table}

Participation at the HealthSoftWare conferences is open to all. Each part is subject to charge, and registration is mandatory. HealthSoftWare decides on the speakers to give a talk or presentation of their experience with the system. Presentations by user groups last 15 to 30 minutes and illustrate how the different sites implement and customise their HealthSoftWare. The presenters provide feedback to the company on what worked well and what to do differently. This also includes concrete requirements that are missing from their point of view such as integration to systems in their infrastructure, or criticism on current design, for example easier reporting functionality.

The HealthSoftWare Conference exists not only of presentations. There are also other forms of events such as discussion rounds, courses and workshops. The user conference that we observed included a day of workshops. The scope of the workshops was defined by HealthSoftWare, and included two topics, i.e. integration of third-party systems with HealthSoftWare platform, and facilitating user customisation of the platform.

In addition to the conference, HealthSoftWare provides courses under the name \textit{HealthSoftWare Academy}. There are several thematic courses with fixed dates that target different user groups or provide in-depth knowledge about certain topics. The courses last one or two days and are subjected to charge. One example of such a course is the \textit{superuser course}, which is for experienced users. The super-user course aims to impart knowledge and skills required to customise the platform by the user organisations for their local context.

HealthSoftWare Academy is organised for a smaller number of participants (compared to the conference) and takes place close to the users' geographical location. While at user conferences the company expects to gain a greater insight into the users' requirements and their adapted solutions, the academy is designed for the users to learn certain skills for using or customising the system. However, as the courses are often "hands-on" and interactive, the organisers also learn from the participants how they use the system and where they might have problems. These insights allow HealthSoftWare designers to improve the platform.

Table~\ref{tab:dimension} summarizes how the four dimensions agenda, participants, scope and resources play out for HealthSoftWare Conferences.
\begin{table}
  \caption{Organisation of user conferences described}
  \label{tab:dimension}
  \begin{tabularx}{0.8\textwidth}{lX}
    \toprule
    Dimension&Findings from Case\\
    \midrule
    Agenda  & Organised by HealthSoftWare. The content of the presentations is decided by the participants. HealthSoftWare decides themes for workshops. \\
    Participants & Participation is open to all. There is a fee for both the user conferences and the academics. The conferences are situated at the location of the company. \\
    Scope & Shaped by HealthSoftWare, through deciding which presenters to invite and what to cover in the workshops. \\
    Resources & Besides the registration fee, the participants need to cover the costs for their travel and stay. Conferences and courses are held during work days, which means that the participants need to be supported by their employers. \\
  \bottomrule
\end{tabularx}
\end{table}

\section{Discussion}

Genuine participation can be defined as \textit{"the fundamental transcendence of the users' role from being merely informants to being legitimate and acknowledged participants in the design process"}~\cite{robertsonParticipatoryDesignIntroduction2012}. Managed communities as a strategy for generification work provide a potentially effective arena for discovering common needs across different user bases. At the same time, these communities legitimatise in a pseudo-democratic way the exclusion of the particular. Requirements relevant only for a few are down-prioritised, and their implementation is either delayed to unforeseen future or outsourced to the users themselves. This is probably the main departure of managed communities from traditional PD processes. Historically, the development of information systems has moved from in-house development to buying generic software~\cite{banslerInformationSystemsDevelopment1994}. When IT is purchased, in form of a generic product or -- in our case -- a platform to be customised, the balance of power is shifted towards a new strong player, i.e. the owner of the product. This implies that managed communities carry multiple, sometimes contradictory roles with respect to PD.

The shift in power is clearly visible with regard to the agenda and the scope of the observed user conference (see e.g. Table~\ref{tab:dimension}). User conferences are normally hosted by a company developing a product with the aim to target a large market. This context influences strongly which information is presented during the conference, and what will become official knowledge, i.e. included in the roadmap for the product. Attention should be also drawn to what is not on the agenda~\cite{starLayersSilenceArenas1999a}, e.g. excess critique or request for odd functionality. The organisers are the actors in power.

There are several ways to "democratise" managed communities. Transferring the concept of managed communities to PD would e.g. imply including the different user groups already at the stage of planning the agenda for the conferences and courses. As user conferences are normally taking place annually, following an iterative approach and reflecting on the previous conference during the planning process is a possible inclusion strategy. Topics for speakers can be decided together with the users -- simple online voting tools can support this. Workshops and courses contribute to the goal of mutual learning, but again the selection process of workshop themes or course topics could be done together with the users and participants. Moreover, to overcome financial and geographic limitations, user conferences can be organised at alternating venues close to the users. For some inspiration for organising more democratic communities one can look at HISP, which has a record of successfully managing a globally distributed user community in a large-scale PD project~\cite{rolandPlatformArchitecturesLargescale2017,braaHealthInformationSystems2012}.

However, we also need to ask for whose benefit such a democratisation process is, and who should pay for it. HealthSoftWare is not a large company. Managing a community is a relatively costly process, which needs to pay off in terms of new sales. It makes sense that the agenda and the scope of these conferences need to fit the product roadmap. Large customers -- such as the municipality who just acquires their self-reporting platform -- and their needs will be prioritised. At the same time, the survival of HealthSoftWare depends in finding new customers. So the conference is also used as a marketing tool. What is guiding a company such as HealthSoftWare is their product roadmap, which needs to be developed based on a sustainable business model for the company.

This raises a fundamental challenge for PD. Traditionally, PD focused on individual user communities interacting with software designers in closed organisational contexts. The focus was on satisfying the needs of \textit{this} community. Even if the system itself was large and complex, its ecosystem was relatively simple. Naturally, platform owners wish to sell their products to multiple user communities. They also want to maximise revenue by segmenting their user base~\cite{pollockGlobalSoftwareIts2007}, and prioritising those users who can pay more and are more aligned with the product roadmap.

Products with ecosystems consisting of numerous communities need a more distributed community management. It is unrealistic to ask one actor in the ecosystem to guarantee value for all other actors through a fully participatory design process. This is valid not only for small companies but also large ones. In studying the ecosystem for Apple's AppStore, Eaton et al.~\cite{eatonDistributedTuningBoundary2015} found that while Apple (similar to HealthSoftWare) played a central role in managing its own user communities, users created communities of their own and used these communities to put pressure on Apple. We have seen similar phenomena within the Norwegian public sector: small municipalities join together when acquiring large IT systems, in this way creating stronger communities managed by themselves and not the vendors.

The concept of managed communities is useful for scaling PD and involving large groups of users and needs to be studied by PD researchers as a potentially new arena for future PD processes.
Our research focused around the user conferences, which are part of a reflection of systems in use. The genuine use of the system in the social environment it is designed for is essential for uncovering the users' need. Further, regular user conferences can support a systematic and iterative design process which acknowledges the emergent use over time.~\cite{heyerDesignEverydayContinuously2010}
However, there are challenges related to (i) how to democratise these communities and the process of creating and maintaining them, and (ii) how to co-manage multiple communities with differing and sometimes conflicting agendas with the goal of creating better IT products. Implementing an iterative and continuous approach to the above two points can support the long-term scaling up of PD projects by establishing structures and routines that can go beyond episodes of participation.

\section{Conclusions}

This paper gives a first glimpse into our investigation of how and if managed communities can support the scaling of participation in PD. The case study is ongoing, and data collection is limited, which can be seen as a limitation. However, we have seen similar patterns across other cases we are currently engaged with. Managed communities have the potential to support more than generifcation work and negotiating user requirements of large communities.

In a next step, we will take the perspective of the users to explore this work further. This case is set in the healthcare sector, hence we will also take into account the voices of patients and indirect caregivers. Although they are not the direct users of the system, they are affected by the system in place and need to be heard.

\bibliographystyle{ACM-Reference-Format}
\bibliography{PDC2020}

\end{document}